# Magnetic Ground State and Spin Excitations in the 2D Trimerized Collinear-II Lattice Antiferromagnet $Li_2Ni_3P_4O_{14}$

**K. S. Chikara, A. K. Bera*, A. Kumar*, and S. M. Yusuf***

*Solid State Physic Division, Bhabha Atomic Research Centre, Mumbai-400085, India*
*Homi Bhabha National Institute, Anusaktinagar, Mumbai-400094, India*

**F. Orlandi[1], and C. Balz[1,2]**

[1]*ISIS Facility, STFC Rutherford Appleton Laboratory, Harwell Oxford, Didcot OX11 0QX, UK*
[2]*Neutron Scattering Division, Oak Ridge National Laboratory, Oak Ridge, Tennessee 37831, USA*

* *Contact author: akbera@barc.gov.in, amitkr@barc.gov.in, smyusuf@barc.gov.in*

**Abstract:**

*We report the magnetic ground state, spin excitations, and spin Hamiltonian of the 2D spin-1 trimerized Heisenberg antiferromagnet $Li_2Ni_3P_4O_{14}$. Below the magnetic ordering temperature $T_N$ = 14.5 K, the compound exhibits a canted long-range antiferromagnetic order with a propagation vector* **k** *= (0 0 0), consistent with the magnetic space group $P2_1/c.1$ (No. 14.75). The ground state magnetic structure consists of ferromagnetic spin-trimers of $Ni^{2+}$ ions. The spin-trimers are coupled antiferromagnetically along the c-axis and ferromagnetically along the a-axis. Inelastic neutron scattering (INS) reveals gapped and dispersive magnon excitations below the $T_N$, and gapless quasi-elastic scatterings at higher temperature. The linear spin-wave theory simulations reveal the essential features of the excitation spectrum; by a spin Hamiltonian composed of ferromagnetic intra-trimer exchange interaction $J_1$ and inter-trimer exchange interactions $J_2$ (FM) and $J_3$(AFM) within the bc plane. The $J_2$ and $J_3$ along the b-axis and c-axis, respectively, with strengths of $J_2/J_1$ =0.79 and $J_3/J_1 \sim = -0.91$. In addition, a weak inter planer ferromagnetic exchange interaction $J_4$ ($|J_4/J_1| \sim 0.12$ is found along the a-axis. The determined exchange constants reveal a 2D trimerized Collinear-II spin lattice within the bc-plane. The analysis of INS spectra by linear spin-wave theory also yields a moderate single-ion anisotropy ($D/J_1$=0.48) which accounts for the observed spin gap below $T_N$ as well as the metamagnetic transition near 44 kOe in dc magnetization (M vs H) curves. These findings identify $Li_2Ni_3P_4O_{14}$ as a rare realization of a two-dimensional trimerized spin system and offer the direct experimental confirmation of theoretically predicted magnon excitations, unveiling the fundamental characteristics of the expected excitation spectrum.*

## I. INTRODUCTION

Understanding quantum many-body interactions in low-dimensional magnetic systems remains a central challenge in modern condensed matter physics [1-4]. Such low-dimensional systems have garnered significant attention in the quantum era, as they give rise to a range of emergent phenomena, including spin-Peierls transitions [5], Bose-Einstein condensation of magnons [6], topological spin textures, such as skyrmions [7], fractionalized spinon excitations in Bethe strings [3], field-induced quantum phase transitions [8], quantum magnetization plateaus and exotic composite quasiparticles, such as doublons and quartrons [9].

In one-dimensional (1D) Heisenberg antiferromagnetic (AFM) uniform chains, the ground state and low-lying excitations strongly depend on the spin value. For spin-1/2 Heisenberg AFM uniform chains, the ground state is a quantum-disordered spin-liquid state characterized by gapless magnetic excitations. In contrast, integer-spin-chains exhibit a gapped singlet ground state, in accordance with the Haldane conjecture. Going beyond uniform spin chains, the introduction of multiple competing exchange interactions leads to spin-cluster formation, giving rise to complex spin geometries, such as zig-zag [10], dimer [11], trimer [12], and tetramer [13] spin chains. Among these, trimerized spin-chain clusters provide an example of a minimally frustrated yet rich platform for studying the interplay of quantum fluctuations and geometrical magnetic frustration. Trimerized spin clusters systems exhibit nontrivial magnetic excitations and serve as ideal candidates to explore emergent quantum phenomena. Recent studies on 1D trimer spin-chains have revealed intriguing quantum mechanical properties, such as quantum phase transitions [14], fractionalization of spin quantum numbers [15], novel composite quasiparticle excitations [16], magnetization plateaus [17], and quantum-spin entanglement [15].

Extending trimerized spin clusters to two-dimensional (2D) lattices introduces new degrees of freedom and novel magnetic behaviors[18]. Recent numerical studies [18] considered different 2D trimerized lattice configurations, such as Collinear-I, Collinear-II, trimerized Lieb, and trimerized hexagon lattices, as shown in Fig. 1, and predicted their ground state magnetic properties and spin excitations. For these 2D trimerized lattices, low-energy excitations are predominantly magnonic for all spin values, unlike their 1D counterparts where spinons dominate for spin-1/2 chains [15]. For weak inter-trimer exchange interactions ($J_2/J_1 < 0.3$, $J_1$ and $J_2$ being the intertrimer and intratrimer exchange interactions), high-energy excitations remain composite in nature, originating from internal trimer dynamics. As inter-trimer exchange coupling strength increases, the composite excitations gradually merge into magnons modes or evolve into continua [18]. Further theoretical studies on trimerized spin-systems indicate that for spin-1 the excitation spectrum bifurcates into low-energy magnons and high-energy composite modes, such as singletons, triplons, pentons, and heptons, which carry effective spin quantum numbers 1, 3, 5, and 7, respectively [19]. These results suggest that spin-1 trimerized 2D systems could serve as fertile ground for realizing unconventional quantum excitations not accessible in spin-1/2 1D or 2D systems.

Several experimental realizations of 1D trimer spin chains have achieved in recent past. For example, $La_4Cu_3Mo_3O_{12}$ [20], $A$$Ni_3(PO_7)_2$ ($A$ = Ca, Sr, Pb) [21], $A_3Cu_3(PO_4)_4$ ($A$ = Ca, Sr, Pb) [22], $SrMn_3P_4O_{14}$ [23], and $Na_2Cu_3Ge_4O_{12}$ [15], have provided valuable insights into the interplay between intra- and inter-trimer exchange interactions on the nature of magnetic ground state and spin excitations including quasiparticle excitations. To date, there have been no experimental reports of a material that exhibits fractionalized quasiparticles and magnon excitations in a two-dimensional trimerized spin lattice. In this context, $Li_2Ni_3P_4O_{14}$ emerges as a potential compound to realize such a 2D trimerized magnetic lattice. The compound crystallizes in a monoclinic crystal structure with space group $P2_1/c$, where a unique arrangement of Ni ions ( Ni1-Ni2-Ni1) forms a spin-1 trimer, and these trimers are coupled via inter-trimer exchange interactions within the *bc* plane [24].

In this work, we present a comprehensive investigation of the magnetic ground state and spin excitations of $Li_2Ni_3P_4O_{14}$ using a combination of experimental techniques, *viz.* neutron powder diffraction, neutron polarization analysis, magnetic susceptibility, specific heat, and inelastic neutron scattering (INS). Below the magnetic ordering temperature ($T_N$) ≈ 14.5 K, $Li_2Ni_3P_4O_{14}$ exhibits commensurate antiferromagnetic order with a propagation vector $k = (0,0,0)$, described by magnetic space group $P2_1/c.1$ (No. 14.75). The INS spectrum reveals gapped spin-wave excitations below the $T_N$ and gapless quasi-elastic excitations above the $T_N$. These low-energy gapped spin-wave excitations are identified as magnons, consistent with theoretical predictions for the trimerized Collinear-II lattice in the strong coupling regime [18]. Through the linear spin-wave theory, we construct the spin Hamiltonian for $Li_2Ni_3P_4O_{14}$, incorporating exchange interactions and single-ion anisotropy. The present study establishes $Li_2Ni_3P_4O_{14}$ as an experimental realization of 2D trimerized spin-1 systems, offering a platform for exploring the unexplored physics of 2D trimerized spin systems.

## II. EXPERIMENTAL METHODOLOGY

Polycrystalline samples of $Li_2Ni_3P_4O_{14}$ were synthesized using a conventional solid-state reaction method in a Muffle furnace. A stoichiometric mixture of NiO (99.99%), $Li_2CO_3$ (99.99%), and $(NH_4)_2HPO_4$ (99.99%) was heated at 1223 K in air for total ~ 170 hours with several intermediate grindings to ensure homogeneity. The phase purity and quality of the sample were verified by x-ray diffraction using a Cu $K\alpha$ radiation.

Bulk magnetization measurements, including DC magnetic susceptibility $\chi_{DC}(T)$, AC magnetic susceptibility $\chi_{AC}(T)$, and isothermal DC magnetization $M(H)$; and specific heat $Cp(T)$, were carried out by using a M/S Cryogenic, UK make Multi Property Measurement System.

One-dimensional neutron-depolarization measurements were carried out using the Polarized Neutron Spectrometer (PNS, $\lambda$ = 1.205 Å) at the Dhruva Research Reactor, BARC, Mumbai, India[25]. The temperature dependence of the transmitted neutron beam polarization was measured in the warming cycle in the presence of a 15 Oe guide field, which preserves neutron beam polarization at the sample position.

Room temperature neutron diffraction pattern was recorded using the powder diffractometer-I (PD-1, $\lambda$ = 1.094 Å) at Dhruva, BARC [26,27] composed of three liner position sensitive $He^3$ neutron detectors covering a momentum transfer range $|Q|$ = 0.5–6.2 $Å^{-1}$ . Low-temperature (over 1.5–100 K) neutron powder diffraction measurements were carried out using the time of flight-based WISH diffractometer at the ISIS Neutron and Muon Source, Rutherford Appleton Laboratory (RAL), UK. The sample was mounted in a vanadium cylindrical can and the temperature was controlled with an oxford instrument cryostat. The diffraction data were diffraction focused in 5 detector banks each covering 32 degrees of the scattering plane [28]. The measured diffraction patterns were analyzed using the Rietveld refinement technique by employing the FULLPROF Suite computer program [29]. The magnetic symmetry of the ordered phase was investigated using the ISODISTORT program [30,31].

The INS measurements were performed on the LET cold neutron time-of-flight spectrometer at the ISIS Neutron and Muon Source, RAL, UK. The LET is equipped with large area position sensitive detectors covering scattering angles from 5° to 140°, enabling measurements over a wide $Q$-range. Approximately 13 gm of powder sample was  packed in a thin aluminium (Al) foil, rolled into annular cylindrical form, and inserted inside a thin-walled cylindrical Al can. The assembly was mounted inside a closed-cycle refrigerator with $^4He$ exchange gas for low-temperature measurements. The rep-rate multiplication method was adopted for  simultaneously collection of data at multiple incident neutron energies ($E_i$ = 22.8, 7.5, 3.7, and 2.2 meV) [32]. The INS spectra were collected at 1.5, 14 and 25 K, with each spectrum measured over 6 hours. The data were processed using the MANTID software package [33]. The spin-wave simulations were carried out using the SPINW program [34], enabling the extraction of the spin Hamiltonian parameters based on the measured experimental INS spectra.

# III. RESULTS

## A. Crystal Structure and Magnetic Phase Transitions:

The crystal structure of $Li_2Ni_3P_4O_{14}$ was investigated using room-temperature neutron powder diffraction [Figure 2 (a)]. The Rietveld refinement of the diffraction data confirms that the compound crystalizes in the monoclinic structure with space group $P2_1/c.1'$ (No. 14). The refined lattice parameters are $a$ = 7.1873(4) Å, $b$ = 7.7493(4) Å, $c$ = 9.3372(5) Å, and $\beta$ = 110.31(4)°, consistent with the previously reported values [24]. The derived fractional atomic coordinates, occupancy factors, and thermal parameters are summarized in Table I. A full site occupancies configuration is found to be consistent with the observed data.

The crystal structure is composed of an edge-sharing $NiO_6$ octahedra containing $Ni^{2+}$ ions ($S$ = 1) [Fig. 2(b)], which form spin trimerized units arranged along the $b$-axis. The special periodicity of the magnetic $Ni^{2+}$ ions, designated as Ni(1) and Ni(2), occupying the crystallographically independent 4$e$ and 2$c$ sites, respectively, leads to the formation of Ni(2)] Ni(1)-Ni(2)-Ni(1) type spin-trimerized

crystal structure. The intratrimer exchange interaction $J_1$ occurs between Ni(1) and Ni(2) through superexchange pathways Ni(1)-O2-Ni(2) along the crystallographic *b*-axis, with a bond length of 3.171(8) Å and bond angles of 94.0(5) and 97.2(5) deg., respectively. These spin trimers are coupled in the *bc* plane through intertrimer exchange interactions $J_2$ and $J_3$ along the *b* and *c* axes, respectively. The $J_2$ interaction along the *b* axis occurs between Ni(1) ions via Ni(1)-O3-Ni(1) and/or Ni(1)-O6-Ni(1) pathways with a Ni(1) − Ni(1) bond length of 3.249(7) Å and bond angle 108.1(6) deg.. The $J_3$ interaction along the *c* axis, between Ni(1) and Ni(2) across trimers (Ni(1) − Ni(2) = 4.694(1) Å), proceeds through extended superexchange networks involving phosphate tetrahedra: Ni(1)-O3- P1-O2-Ni(2) and Ni(1)-O5-P2-O6-Ni(2). A possible weak interlayer exchange interaction ($J_4$) along the *a*-axis (Ni(1)–Ni(1) = 5.043(3) Å) may also occur via Ni(1)-O3- P1-O2-Ni(1)/ Ni(1)-O5- P2-O6-Ni(1) pathways. Although $J_2$ and $J_3$ exchange interactions occur through different geometries, their effective strengths are found to be comparable (discuss later), whereas $J_4$ is considerably weaker as established from INS data. This arrangement supports a 2D trimerized lattice topology of either collinear-I or collinear-II type, as discussed in detail in magnetic excitations part.

The macroscopic magnetic properties of $Li_2Ni_3P_4O_{14}$ were investigated through temperature- and field-dependent DC magnetization measurements. The zero-field-cooled (ZFC) and field-cooled warming (FCW) DC-susceptibility [$\chi_{DC}(T)$] curves, measured under an external field of 100 Oe, are shown in Fig. 3(a). An increase in susceptibility below 14.5 K signals the onset of a long-range magnetic order. This magnetic phase transition is corroborated by anomalies in the specific heat and AC susceptibility [$\chi_{ac}(T)$] [Figs. 3(b) and (c)]. A λ-type peak in $C_p(T)$ curve under zero magnetic field confirms the development of an AFM-like ground state, similar to that observed for the isostructural $A$$Ni_3P_4O_{14}$ ($A$ = Ca, Sr, Pb) compounds [21]. The field-dependent magnetization, measured at 5 K [Fig. 3(e)], shows a linear increase with a subtle anomaly near 44 kOe, indicative of a metamagnetic transition. Notably, the magnetization at 90 kOe reaches only ~0.9 $\mu_B/Ni^{2+}$, well below the expected saturation value of ~2 $\mu_B/Ni^{2+}$, implying that significantly high magnetic fields are required to achieve the saturation. Here we would like to mentioned that 1D spin-1 trimer chains are theoretically predicted to exhibit quantized magnetization plateaus at 1/3 and 2/3 of the saturation value, i.e. at 0.66 $\mu_B/Ni^{2+}$ and 1.33 $\mu_B/Ni^{2+}$, respectively, in accordance with the Oshikawa's condition. These plateaus are expected to show concave curvature at low field values due to low-lying quantum excitations [9]. In contrast, the observed linear M vs H curve and the absence of any magnetization plateau indicate towards a 2D trimerized lattice, for which magnetization plateau states have not been theoretically predicted. Additional support for the onset of long-range order comes from neutron depolarization measurements, performed under a magnetic field of 15 Oe [Fig. 3(d)]. A clear depolarization of the transmitted neutron beam below 14.5 K indicates the presence of a finite internal domain magnetization, consistent with the $\chi_{DC}(T)$ behaviour in Fig.3(a) and a weak hysteresis seen in $M(H)$ curve at 5 K in Fig. 3(e). The finite domain magnetization could arise from either a ferrimagnetic component of the long-range order, and/or weak ferromagnetic components and/or incomplete cancellation of antiferromagnetically aligned moments. The microscopic origin of this spontaneous magnetization and the detailed nature of the magnetic ground state will be elucidated through neutron scattering studies, which are presented in the following sections.

## B. Microscopic magnetic properties

### 1. Magnetic ground state:

The microscopic nature of the magnetic ground state in $Li_2Ni_3P_4O_{14}$ has been investigated by temperature dependent neutron powder diffraction measurements over the temperature range 1.5- 100 K [Fig. 4]. The pure magnetic diffraction pattern is obtained (shown in the top panel of Fig. 4) after subtraction of diffraction pattern measured at 16 K (paramagnetic state) from low temperature (1.5 K) diffraction pattern. The neutron powder diffraction patterns measured at 100 K (paramagnetic state) and 1.5 K (magnetically ordered state) are shown in Fig. 5 along with calculated patterns. The experimental pattern at 100 K could be well reproduced using the monoclinic space group $P2_1/c.1'$ (No. 14), as found at room temperature neutron diffraction study [Fig. 2]. Minor traces of $Li_4P_2O_7$ (~2.55 wt%) and NiO (~0.92 wt%) were detected as secondary phases. Nevertheless, these secondary phases do not hinder determination of the magnetic structure of the main phase as $Li_4P_2O_7$ is non-magnetic and contribution from NiO is negligible due to low concentration. Below the magnetic transition temperature ($T_N$ = 14.5 K), additional Bragg peaks appear [Fig. 5(b)], which are forbidden under the $P2_1/c.1'$ parent space group symmetry, confirming the onset of a long-range antiferromagnetic ordering. All magnetic Bragg peaks could be indexed using a commensurate propagation vector $k$ = (0 0 0) indicating the same periodicity of the crystallographic and magnetic unit cells. Symmetry analyses were performed using the ISODISTORT program [30,31] to determine the symmetry-allowed magnetic structures and corresponding magnetic space groups by considering both Ni sites (Wyckoff positions 4*e* and 2*c*) as magnetically active. The analysis yielded four irreducible representations (IRs): $m\Gamma1^+$, $m\Gamma1^-$, $m\Gamma2^+$, $m\Gamma2^-$ [presented in Table II], corresponding to four one dimensional magnetic space groups (Shubnikov Groups) in Belov-Neronova-Smirnova (BNS) [36] notation. Here we would like to mention that all the IRs transform from parent space group with following transformations *a*, *b*, *c*; 0, 0, 0. Only the IRs $m\Gamma1^+$, and $m\Gamma2^+$ allow magnetic moments at both Ni sites, whereas, the IRs $m\Gamma1^-$, and $m\Gamma2^-$ prohibit moment formation on the 2*c* site to preserve the symmetry. The direction of the magnetic moment on a given site within a unit cell is constraint by the symmetry of the magnetic space group. According to the Landau theory, a second order magnetic phase transition involves a single irreducible representation. Since, there is only one magnetic phase transition in $Li_2Ni_3P_4O_{14}$ at 14.5 K, the magnetic ground state must correspond to either $m\Gamma1^+$ or $m\Gamma2^+$. Both the IRs allow moment components along all three axes (*i.e* $m_x$, $m_y$, and $m_z$). The $m\Gamma1^+$ allows ferromagnetic (FM) coupling along the *b*-axis and AFM alignment along the *a* and *c*-axes, whereas, $m\Gamma2^+$ permits AFM order along the *b* axis, and FM along the *a* and *c* axes [Table III]. The Rietveld analyses of the 1.5 K pattern revealed that $m\Gamma1^+$ representation reproduces the experimental magnetic diffraction pattern, whereas, $m\Gamma2^+$ could not reproduce all the magnetic peak intensities. Thus, the magnetic structure of $Li_2Ni_3P_4O_{14}$ is best described by the Shubnikov magnetic space group $P2_1/c.1$ (No. 14.75 corresponding to $m\Gamma1^+$) by considering the same unit cell as that of the paramagnetic phase. The fitted pattern at 1.5 K is shown in Fig. 5(b). The magnetic structure, depicted in Fig. 6(a), consists of FM Ni(1)–Ni(2)–Ni(1) trimers that are coupled ferromagnetically along the *b*-axis, antiferromagnetically along the *c*-axis, and ferromagnetically along the *a*-axis. Overall, this results in a canted antiferromagnetic ground state, stabilized by intra- and inter-trimer exchange interactions. The derived magnetic moment components at 1.5 K are: $m_x$ = -0.60 ± 0.01 $\mu_B$, $m_y$ = -0.48 ± 0.04 $\mu_B$, and $m_z$ = 1.62 ±0.02 $\mu_B$. The moments predominantly align along the *c*-axis. Importantly, the symmetry of $P2_1/c.1$ allows a FM alignment along the *b*-axis for both Ni(1) and Ni(2), which results in a net uncompensated moment of ~ 0.16 $\mu_B/Ni^{2+}$ per unit cell. Such a FM component explains experimentally observed weak hysteresis in $M(H)$ curves as well as finite neutron beam depolarization below the $T_N$ = 14.5 K. The derived total order moment value, 1.97± 0.04 $\mu_B$, is in good agreement with the spin only value (2$\mu_B$) for $Ni^{2+}$ ($S$ = 1). The variation of the ordered site moment with the temperature is shown in Fig. 6(b). To gain further insight into the origin of the determined magnetic ground state, the strength and nature of the exchange interactions, inelastic neutron scattering measurements were employed and discuss in the next section.

### 2. Magnetic Excitations and Spin Hamiltonian:

Inelastic neutron scattering (INS) is a powerful technique for probing magnetic excitations and determining the spin Hamiltonian of magnetic materials. The INS spectra of $Li_2Ni_3P_4O_{14}$, measured at $T$ = 1.5, 14 and 25 K with various incident neutron energies ($E_i$ = 22.8, 7.5, and 2.2 meV), are shown in Fig. 7 as two-dimensional ($Q$, $E$) intensity maps. Below the $T_N$ = 14.5 K, the spectra reveal three broad excitation bands in ~1–2.5, ~2.8–4.6, and ~5.3–6.5 meV range, indicating the presence of well-defined spin-wave excitations [Figs. 7 and 8]. The magnetic origin of these scattering modes is evident from the decreasing intensity with increasing momentum transfer $|Q|$ as well as increasing temperature. Moreover, no phonon-like modes are observed across the measured lower ($Q$, $E$) space for $E_i$=7.5 meV, as evident from no observed enhancement of inelastic scattering with increasing temperature or increasing $|Q|$. This results in clean magnetic spectra without the need for any phonon background subtraction. The INS spectra measured with higher incident energy of $E_i$=22.8 meV [ Figure 7(g-i)] reveal very weak phonon scatterings in the region of high momentum transfer ($Q$ > 4 Å$^{-1}$) and high energy transfer ($E$ > ~7.5 meV). Although directional information is lost in powder INS spectra due to powder averaging, energy dependent singularities remain present in the density of states which are generally sufficient to determine the spin Hamiltonian (the strength and sign of exchange interactions and anisotropy parameters). At low temperatures, the excitation spectrum exhibits a finite energy gap. As temperature increases, these excitation energy modes soften, and excitation spectra become gapless and quasielastic above the $T_N$. This implies that the low-|$Q$| scattering is related to the spin-wave excitations in the canted AFM ground state of the material. Similar excitation modes are observed in the isostructural compound $CaNi_3P_4O_{14}$ [37]. To quantitatively interpret the experimentally observed magnetic spectrum of $Li_2Ni_3P_4O_{14}$, we have simulated powder average spin-excitation spectra using the SPINW program and have compared with the experimental spectra. The magnetic lattice contains only $Ni^{2+}$ ($3d^8$, $S$ = 1) magnetic ions, forming trimerized spin units within the two-dimensional network in the $bc$ plane. The model includes four distinct Heisenberg exchange couplings, *viz.* $J_1$, $J_2$, $J_3$, and $J_4$, corresponding to a trimerized Collinear-II lattice [Figs. 1 and 2]. To account for the observed excitation gap [Fig. 7], a single-ion anisotropy ($D$) term was included in the Hamiltonian, attributed to the local crystal field effects on the $Ni^{2+}$ ions in a distorted $NiO_6$ octahedron. For simplicity, the same $D$ value was considered for both the Ni(1) and Ni(2) sites. The spin Hamiltonian for the present trimerized Collinear-II lattice is expressed as:

$$H = \sum_{k=1}^{N/3}[J_1(\vec{S}_{k,1}.\vec{S}_{k,2} + \vec{S}_{k,2}.\vec{S}_{k,3}) + J_2(\vec{S}_{k,3}.\vec{S}_{k+1,1})] + \sum_{k,l} J_3(\vec{S}_k.\vec{S}_l) + \sum_{k,l'} J_4(\vec{S}_k.\vec{S}_{l'}) + \sum_k D(S_k^z)^2 \quad (1)$$

where $\vec{S}_{k,i}$ is a spin-1 operator at $i^{th}$ (=1,2, and 3) site of $k^{th}$ spin-trimer along $b$-axis and term $\vec{S}_k.\vec{S}_l$ is interaction between $k^{th}$ trimer and $l^{th}$ trimer along the $c/a$ axis. $J_1$ is the intra-trimer exchange interaction, $J_2$, $J_3$, and $J_4$ are inter-trimer exchange interactions along the $b$, $c$, and $a$-axes, respectively [Fig. 2(b)]. In our simulations, only nearest-neighbor superexchange interactions were considered and long-range exchange interactions were neglected due to the localized ionic nature of Ni 3$d$ orbitals. The magnetic ground state configuration, as obtained from the neutron diffraction study, was used for the simulation of the spin-wave spectra. By systematically adjusting the exchange parameters ($J_1$, $J_2$, $J_3$, and $J_4$) and anisotropy $D$, a good solution was found for $J_1$= -0.83 meV (FM), $J_2$= -0.66 meV (FM), $J_3$= 0.75 meV (AFM), $J_4$ = -0.10 (FM) and $D$ = -0.40 meV. The simulated spin-wave intensity map, shown in Figs. 9 (a) and (b), successfully reproduces the main features of the experimental INS spectra. The comparison between experimental and simulated spectra for $E_i$ = 7.5 meV is shown in Figs. 9 (a) and (b). Further comparisons of constant-energy and constant-$Q$ cuts are shown in Figs. 9(c) and (d), respectively, demonstrating good agreement between experimental and theoretical spectra. The signs of the exchange constants i.e, FM $J_1$, FM $J_2$, AFM $J_3$ and FM $J_4$ are compatible with the magnetic ground state (established from the neutron diffraction analysis) i.e., FM trimers which are aligned ferromagnetically along the $b$-axis, antiferromagnetically along the $c$-axis and ferromagnetically along the $a$-axis, respectively.

## IV. DISCUSSION

The topology of magnetic ions in a crystal lattice can be trimerized in various configurations, either one-dimensional or two-dimensional, each influencing the nature of magnetic excitations in distinct ways as discussed in introduction section. The experimental validation of the 2D trimerized model remains an open question, though recently numerical studies have provided valuable predictions regarding their excitation spectra. Theoretical calculations suggest the emergence of composite quasiparticles, such as doublons and quatrons for spin-1/2, and singletons, triplons, pentons, heptons, etc. for spin-1 systems in the weak intertrimer ($J_2/J_1$< 0.3) interaction regime. However, the dispersion relations of the composite excitations strongly depend on the underlying lattice geometry, distinguishing Collinear-I, Collinear-II, Lieb, and hexagonal lattices. In 2D trimerized models, the low-energy excitations are magnon-like for all spin values in contrast to 1D trimer chains having two-spinon continua for spin-1/2 and magnon excitations for spin-1 systems. The numerical calculations for 2D lattices further reveal that as the parameter $J_2/J_1$ gradually increases beyond 0.3, the quasi-particle excitations above magnons modes begin to hybridize with the low-energy magnons, leading to the formation of a broad continuum. When $J_2/J_1 = 0.7$, all the excitations merge into single magnon-like mode accompanied by a weak residual continuum. In the strong inter-trimer coupling regime ($J_2/J_1 > 0.7$), the low energy magnon excitations are well-described by the semiclassical linear spin wave theory (LWST), indicating the development of a long-range magnetic order, as observed for the studied compound $Li_2Ni_3P_4O_{14}$ [Figs. 7 and 9]. The compound $Li_2Ni_3P_4O_{14}$ exhibits a trimerized spin structure consistent with the Collinear-II lattice model. The key difference between collinear-I and a collinear-II lattices lies in the number and arrangement of inter-trimer couplings. In a collinear-I lattice, each trimer connects to adjacent trimers via three vertical inter-trimer exchange interactions [Fig. 1]. In the case of collinear-II lattice, only one vertical inter-trimer coupling connects adjacent trimer [Fig. 1]. Our LSWT analysis confirms that $Li_2Ni_3P_4O_{14}$ is best described by the Collinear-II lattice, with one dominant vertical inter-trimer coupling (antiferromagnetic $J_3$, along the *c*-axis). In addition, a ferromagnetic inter-trimer coupling $J_2$ is present along the *b*-axis. Both inter-trimer exchange interactions are found to be significant ($J_2/J_1 = 0.79$, $J_3/J_1 = 0.91$). This place the present compound $Li_2Ni_3P_4O_{14}$ in the strong inter-trimer coupling regime ($J_2/J_1 > 0.7$) of the trimerized Collinear-II lattice model, where a dominant magnon mode in the excitation spectrum is theoretically predicted.

Now we compare $Li_2Ni_3P_4O_{14}$ with other members of the $ANi_3P_4O_{14}$ (*A* = Ca, Sr, Pb) family [21] to understand the effects of the A-site substitution on structural, magnetic, and spin excitation properties. The substitution of monovalent Li ions in place of larger divalent cations significantly alters the crystal structural parameters, specially reduction in *a*, *c*, and $\beta$, while *b* remains nearly unaffected [see Table IV, Fig. 10]. These structural changes directly influence the nature and strength of the exchange interactions, as revealed by our INS study. For example, the Ca-based compound [37] exhibits stronger exchange interactions ($J_1$= 1.30 meV, $J_2/J_1 \sim 0.81$, ($|J_3/J_1| \sim 0.69$)), and a weaker single-ion anisotropy ($D = 0.25$ meV) compared to the Li-based compound. This trend indicates that the smaller $Li^+$ ion leads to enhanced anisotropy and weaker exchange interactions, possibly due to lattice contraction and local distortions of the $NiO_6$ octahedra. Further evidence of these effects can be seen in the evolution of the $T_N$ and the critical field for the metamagnetic transition ($H_{SF}$) across the $ANi_3P_4O_{14}$ series [Fig. 10]. The AFM ordering temperature ($T_N$) increases initially with increasing A-site ionic radius but decreases beyond a certain point, while $H_{SF}$ displays an inverse correlation. Notably, compounds with lower $T_N$ value tend to exhibit higher critical field value, reflecting a stronger single-ion anisotropy. Consistent with this trend, $Li_2Ni_3P_4O_{14}$ exhibits a relatively high $H_{SF}$ valueand a substantial anisotropy gap in its excitation spectrum. As temperature approaches $T_N$, the excitation gap softens and eventually closes, indicating that the anisotropy is intrinsically linked to the long-range magnetic order in the ground state.

## V. SUMMARY AND CONCLUSIONS

The macroscopic and microscopic magnetic properties of the 2D spin-trimer compound $Li_2Ni_3P_4O_{14}$, featuring collinear-II lattice, have been thoroughly investigated using temperature- and field-dependent bulk magnetization, along with low-temperature neutron elastic and inelastic scattering measurements. The compound undergoes a magnetic phase transition into a long-range canted antiferromagnetic (AFM) ground state below $T_N = 14.5$ K, characterized by the propagation vector $\boldsymbol{k} = (000)$ and magnetic space group $P2_1/c.1$ (No. 14.75). The magnetic ground state consists of ferromagnetic (FM) trimers of $Ni^{2+}$ ($S = 1$). Such trimers are ferromagnetically coupled along the *b*-axis and antiferromagnetically coupled along the *c*-axis within the *bc*-plane with an ordered moment value of ~ 1.97 $\mu_B/Ni^{2+}$. Antiferromagnetic alignment of spin components ($m_x = -0.60 \pm 0.01\ \mu_B$ and $m_z = 1.62 \pm 0.02\ \mu_B$) along the *a*- and *c*-axes, respectively, result into a net zero magnetization within an unit cell, whereas, the ferromagnetic alignment of spin components ($m_y = -0.48 \pm 0.04\ \mu_B$) along the *b*-axis result in a net magnetization of ~ 0.16 $\mu_B/Ni^{2+}$ per unit cell. Inelastic neutron scattering measurements below $T_N$ reveal gapped and dispersive spin-wave excitations, consistent with a spin-anisotropic magnetic ground state. Detailed modelling of the excitation spectra using the linear spin-wave theory reveals that the system is best described as a 2D trimerized collinear-II lattice, with comparable strengths of inter-trimer exchange interactions $J_2$ (-0.66 meV) and $J_3$ (0.75 meV), and dominant intra-trimer ferromagnetic exchange $J_1$ (-0.83 meV). These findings highlight the importance of trimer topology and anisotropy in shaping the excitation spectrum. Our results provide a comprehensive picture of the magnetic behavior in $Li_2Ni_3P_4O_{14}$ and establish it as a model system for 2D spin-1 trimerized magnets. The observed strong intertrimer couplings place the system in the regime where a single magnon mode dominates, in agreement with theoretical predictions for 2D collinear-II trimerized spin lattices. This study paves the way for exploring quasiparticle excitations in related 2D trimerized systems with weaker intertrimer coupling, and contributes to the broader understanding of quantum magnetism in low-dimensional spin-cluster systems.

**ACKNOWLEDGMENTS:**

Authors thank the Department of Science and Technology (DST), India (SR/NM/Z-07/2015) for the neutron scattering experiment through Indian Access beamtime, and Jawaharlal Nehru Centre for Advanced Scientific Research (JNCASR) for managing the project. Experiments at the ISIS Neutron and Muon Source were supported by beamtime allocations RB2310336 and RB2368052 from the Science and Technology Facilities Council. SMY acknowledges the financial assistance from ANRF, DST, Govt. of India, under the J. C. Bose fellowship program (JCB/2023/000014).

## TABLES:

**TABLE I.** Rietveld-refined fractional atomic coordinates, isotropic thermal parameters and site occupancies for $Li_2Ni_3P_4O_{14}$ at room temperature

| **Atom** | **Site** | ***x/a*** | ***y/b*** | ***z/c*** | $B_{iso}$ | ***Occ.*** |
|---|---|---|---|---|---|---|
| Li | 4*e* | 0.6921 (5) | 0.0851(4) | 0.6464(3) | 0.65(5) | 1.0 |
| Ni1 | 4*e* | 0.8088 (8) | 0.1314(9) | 0.9780(6) | 0.40(1) | 1.0 |
| Ni2 | 2*c* | 0.0 | 0.0 | 0.5 | 0.40(4) | 1.0 |
| P1 | 4*e* | 0.1158(1) | 0.2145(7) | 0.8236(4) | 0.29(5) | 1.0 |
| P2 | 4*e* | 0.4031(1) | 0.9594 (1) | 0.7855(14) | 0.34(9) | 1.0 |
| O1 | 4*e* | 0.3214(14) | 0.1287(5) | 0.8100(10) | 0.58(2) | 1.0 |
| O2 | 4*e* | 0.0311(14) | 0.0801(4) | 0.8996(12) | 0.55(5) | 1.0 |
| O3 | 4*e* | 0.9882 (5) | 0.2074(13) | 0.6457(13) | 0.63(8) | 1.0 |
| O4 | 4*e* | 0.6198(8) | 0.9854(6) | 0.8034(10) | 0.62(8) | 1.0 |
| O5 | 4*e* | 0.3900(15) | 0.8243(5) | 0.9061(13) | 0.77(3) | 1.0 |
| O6 | 4*e* | 0.2713(9) | 0.8881(5) | 0.6285(11) | 0.79(3) | 1.0 |
| O7 | 4*e* | 0.1674(5) | 0.3792(7) | 0.8960(11) | 0.53(9) | 1.0 |

**TABLE II.** Magnetic space groups generated using the ISODISTORT program based on the crystallographic space group $P2_1/c.1'$ and propagation vector $k$ = (0 0 0) for $Li_2Ni_3P_4O_{14}$

| **IR** | **Order parameter direction** | **Magnetic Space Group (BNS setting)** | **M. S.G. Number** | **Wykoff position** | **Symmetry constraints on moment** |
|---|---|---|---|---|---|
| mΓ1+ | (*a*) | $P2_1/c.1$ | 14.75 | 4*e*(Ni1) | $(m_x, m_y, m_z)$ |
| | | | | 2*c*(Ni2) | $(m_x, m_y, m_z)$ |
| mΓ2- | (*a*) | $P2_1'/c$ | 14.77 | 4*e*(Ni1) | $(m_x, m_y, m_z)$ |
| | | | | 2*c*(Ni2) | (0, 0, 0) |
| mΓ1- | (*a*) | $P2_1/c'$ | 14.78 | 4*e*(Ni1) | $(m_x, m_y, m_z)$ |
| | | | | 2*c*(Ni2) | (0, 0, 0) |
| mΓ2+ | (*a*) | $P2_1'/c'$ | 14.79 | 4*e*(Ni1) | $(m_x, m_y, m_z)$ |
| | | | | 2*c*(Ni2) | $(m_x, m_y, m_z)$ |

**TABLE III.** Possible magnetic moment components at magnetic Wykoff positions for the allowed magnetic space groups of $Li_2Ni_3P_4O_{14}$

| Magnetic Space Group | Wykoff Positions | |
|---|---|---|
| | 4*e*(Ni1) | 2*c*(Ni2) |
| $P2_1/c.1$ | $(x, y, z \mid m_x, m_y, m_z)$ | $(0, 0, 1/2 \mid m_x, m_y, m_z)$ |
| | $(-x, y+1/2, -z+1/2 \mid -m_x, m_y, -m_z)$ | $(0, 1/2, 0 \mid -m_x, m_y, -m_z)$ |
| | $(-x, -y, -z \mid m_x, m_y, m_z)$ | |
| | $(x, -y+1/2, z+1/2 \mid -m_x, m_y, -m_z)$ | |
| $P2_1'/c'$ | $(x, y, z \mid m_x, m_y, m_z)$ | $(0, 0, 1/2 \mid m_x, m_y, m_z)$ |
| | $(-x, -y, -z \mid m_x, m_y, m_z)$ | $(0, 1/2, 0 \mid m_x, -m_y, m_z)$ |
| | $\mathbf{(-x, y+1/2, -z+1/2 \mid m_x, -m_y, m_z)}$ | |
| | $\mathbf{(x, -y+1/2, z+1/2 \mid m_x, -m_y, m_z)}$ | |

**TABLE IV.** Ion radii of *A*-site ions, lattice parameters, Néel temperatures ($T_N$), critical fields for spin-flop metamagnetic transition ($H_{SF}$), Ni-Ni distances, and Ni-O-Ni bond angles for $ANi_3P_4O_{14}$ (*A* = Li, Ca, Sr, Pb, and Ba)

| | | Li **[Present]** | Ca[21] | Sr[21] | Pb[21] | Ba[21] |
|---|---|---|---|---|---|---|
| | Ion radius (Å) | 0.59 | 0.99 | 1.12 | 1.20 | 1.34 |
| | *a* (Å) | 7.1873(4) | 7.330 | 7.407 | 7.402 | 7.526 |
| | *b* (Å) | 7.7493(4) | 7.589 | 7.658 | 7.673 | 7.779 |
| | *c* (Å) | 9.3372(5) | 9.398 | 9.440 | 9.468 | 9.573 |
| | $\beta$ (deg.) | 110.31(4) | 112.1 | 112.2 | 112.4 | 112.8 |
| | $T_N$ (K) | 14.5 | 15.8 | 15.3 | 15.6 | 14.5 |
| | $H_{SF}$ (T) | 4.4 @5K | 2.2 @2K | 3.5 @2K | 3.9 @2K | 5.0 @2K |
| Along the *b* direction | Ni(1)-Ni(2) | 3.171(8) | 3.11 | 3.14 | 3.15 | 3.19 |
| | Ni-O-Ni (deg) | 94.0(5) | 93.6 | 93.5 | 93.8 | 93.9 |
| | | 97.2(5) | 98.9 | 100.6 | 100.7 | 101.0 |
| Along the *b* direction | Ni(1)-Ni(1) (Å) | 3.249(7) | 3.21 | 3.23 | 3.22 | 3.27 |
| | Ni-O-Ni (deg) | 108.1(6) | 102.6 | 102.7 | 102.4 | 102.7 |
| Along the *c* direction | Ni(1)-Ni(2) (Å) | 4.694(1) | 4.70 | 4.72 | 4.73 | 4.77 |

# FIGURES:

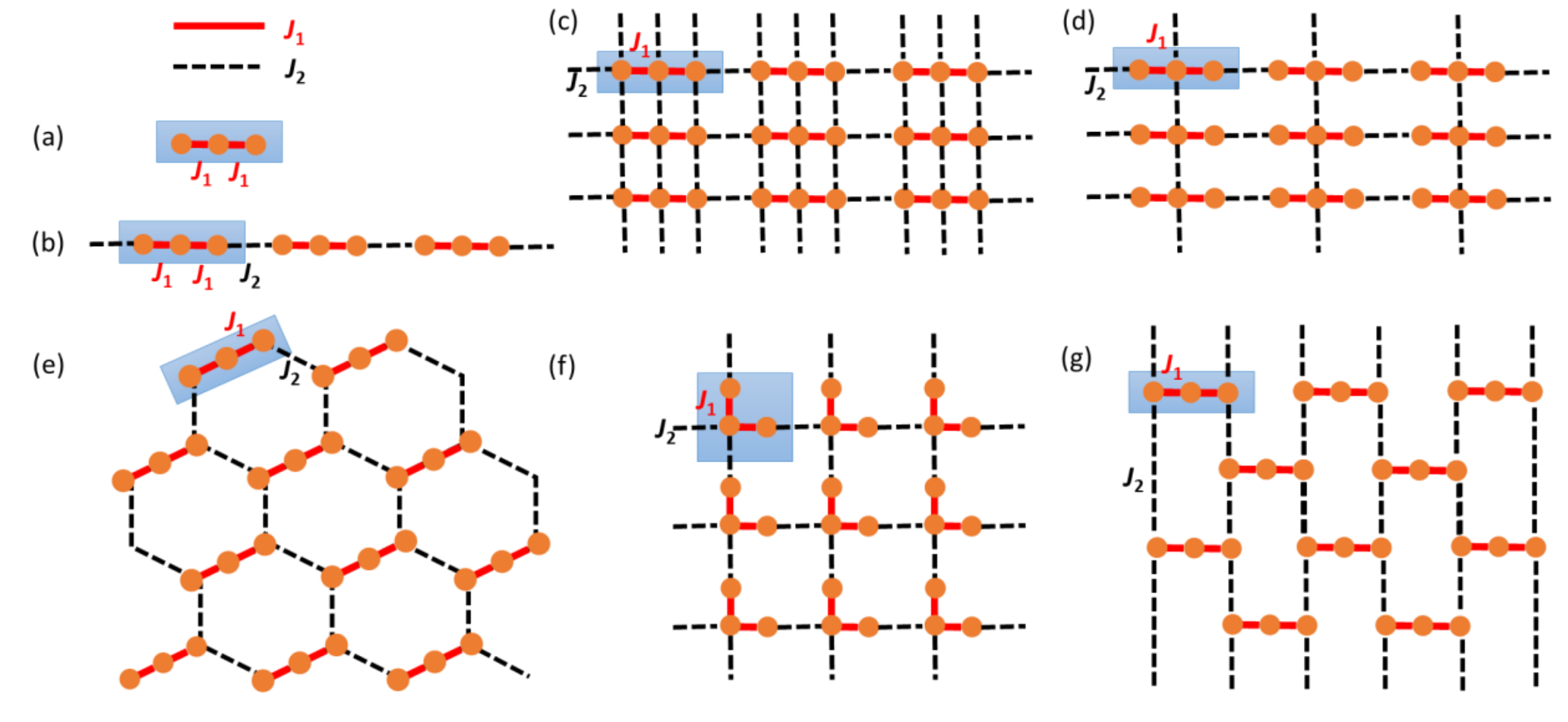

**FIG. 1.** (a) Schematics of (a) an isolated trimer cluster, (b) a linear trimer chain, (c) a collinear-I lattice, (d) a collinear-II lattice, (e) a trimerized hexagonal lattice, (f) a trimerized Lieb lattice, and (g) a topological equivalent lattice of Hexagon lattice. Solid circles represent the magnetic ions, $J_1$ is the intratrimer exchange interaction, and $J_2$ is the intertrimer exchange interaction.

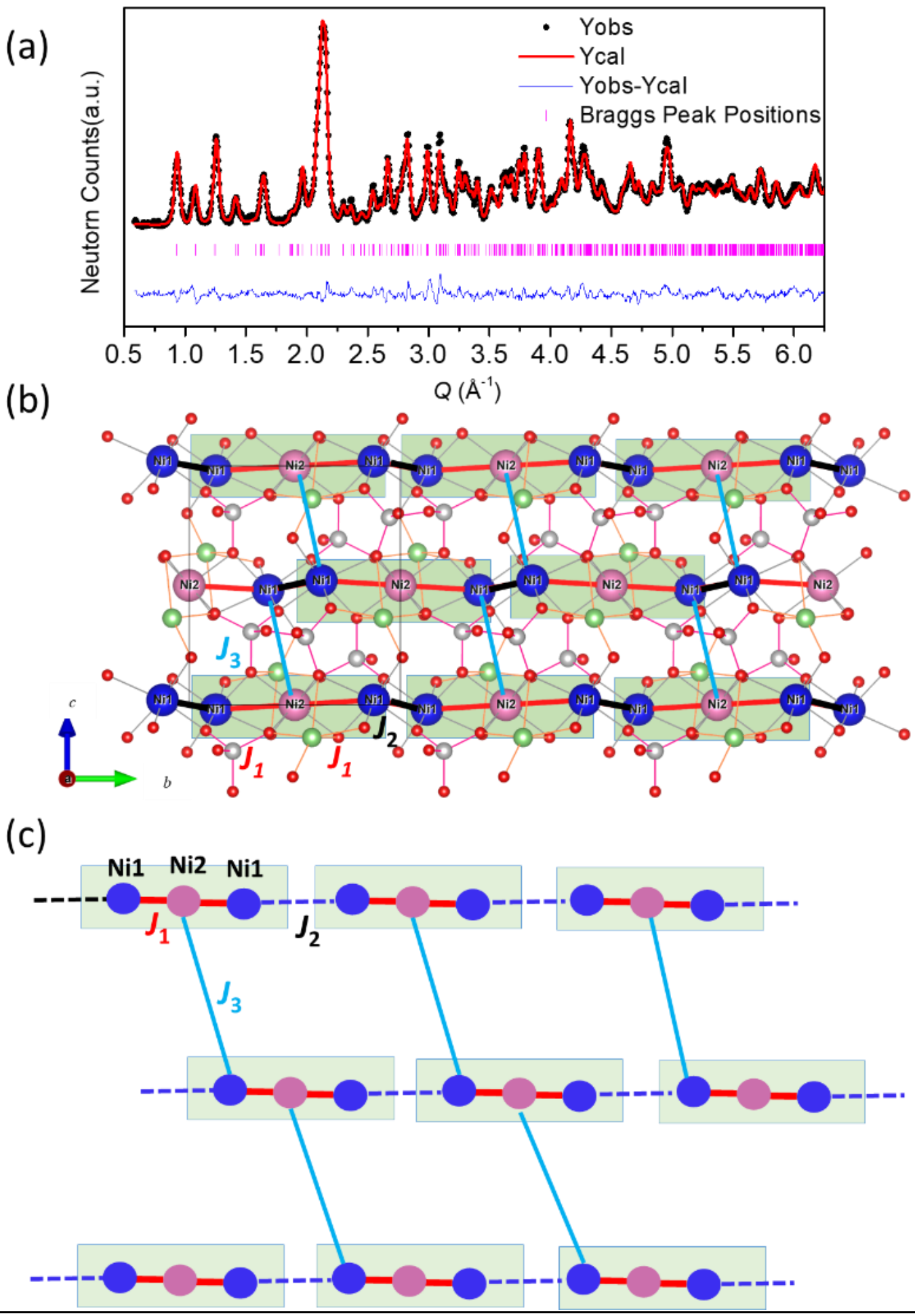

**FIG. 2.** (a) Experimentally observed (circles) (measured at 300 K using PD-1 diffractometer) and calculated (solid line) neutron diffraction patterns of $Li_2Ni_3P_4O_{14}$ .. The difference between observed and calculated patterns is shown by the solid blue line at the bottom. The vertical bars indicate the positions of allowed nuclear Bragg peaks. (b) Schematic representation of crystal structure and exchange interactions in $Li_2Ni_3P_4O_{14}$. The Ni(1), Ni(2), P, Li and O ions are shown as blue, pink, grey, green, and red spheres, respectively. Intratrimer exchange interactions along the *b*-axis between Ni(1) and Ni(2) are denoted by $J_1$, whereas intertrimer interactions between Ni(1) atoms along the *b*-axis are denoted by $J_2$. The intertrimer exchange interaction along the *c*-axis between the nearest Ni(1) and Ni(2) is labeled as $J_3$. The individual trimer units are enclosed in blue ellipsoids. (c) Simplified schematic illustrating the magnetic exchange pathways in the same *bc*-plan as shown in (b).

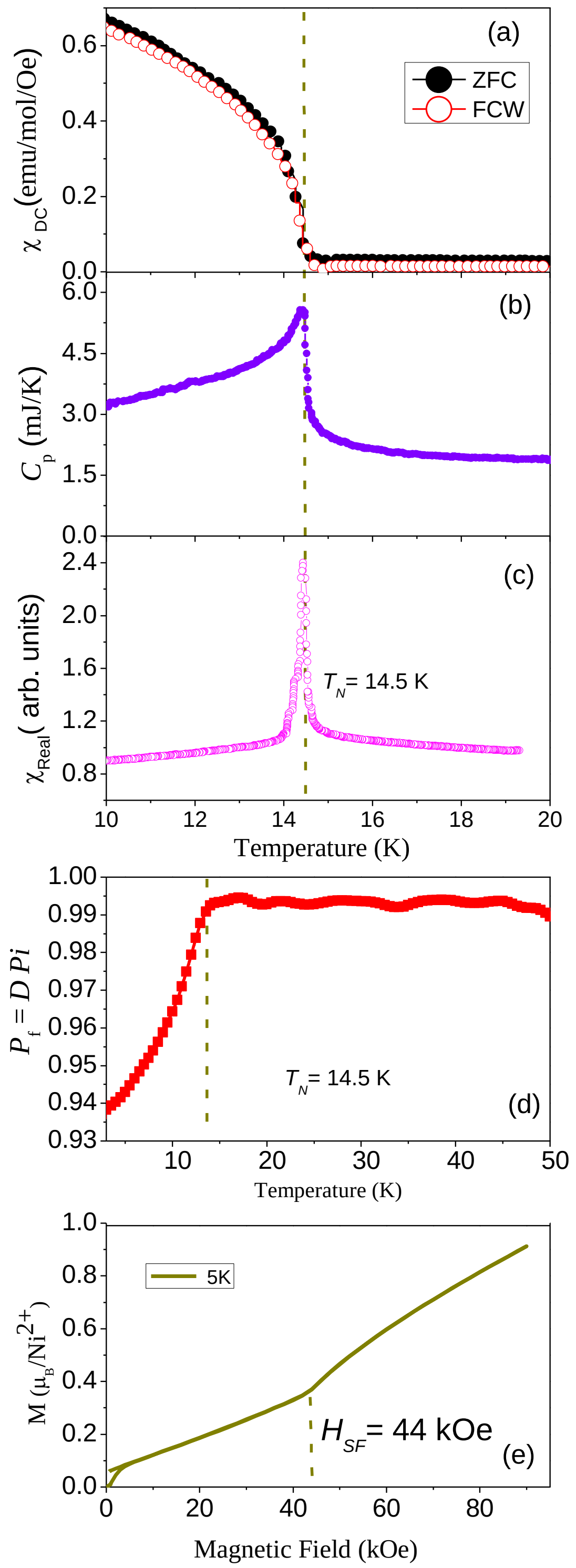

**FIG. 3.** Temperature dependence of (a) DC magnetic susceptibility ($\chi_{DC} = M/H$) measured under an applied magnetic field of 100 Oe using the ZFC and FCW protocols; (b) specific heat; and (c) ac susceptibility curves, respectively. The magnetic ordering temperature is marked with a vertical line. (d) Transmitted neutron beam polarization ($P_f$) measured as a function of temperature under 15 Oe magnetic guide field. (e) Magnetic field dependence of DC magnetization at 5 K, revealing a metamagnetic transition at 44 kOe.

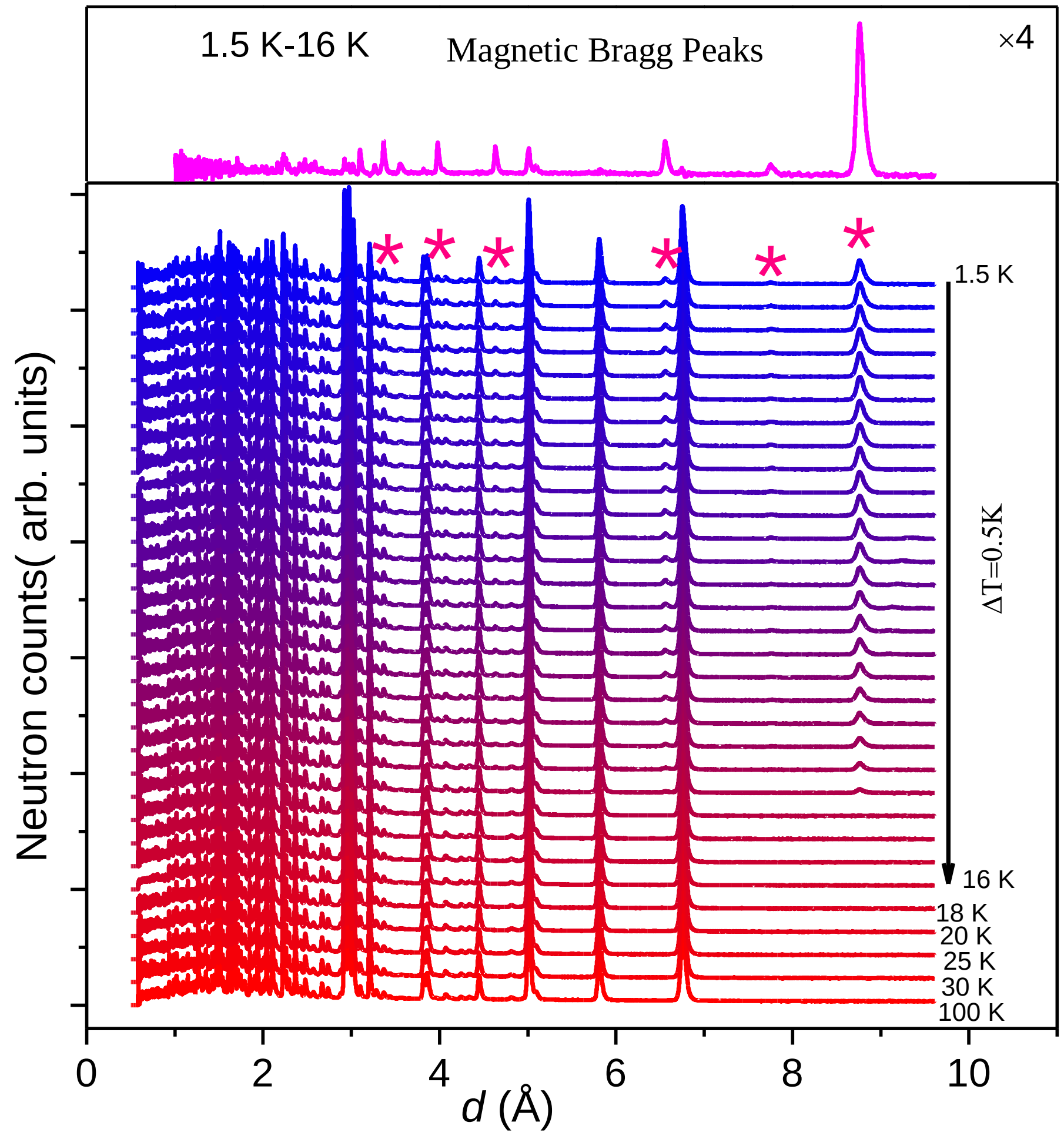

**FIG. 4.** Low-temperature neutron diffraction patterns of $Li_2Ni_3P_4O_{14}$ measured using the WISH diffractometer at ISIS, RAL, UK. The top panel shows the (1.5 K – 16 K) difference pattern highlighting magnetic Bragg peaks at 1.5 K. The intensity of magnetic pattern in the top panel is zoomed 4 time for better visualization of the magnetic Braggs peak.

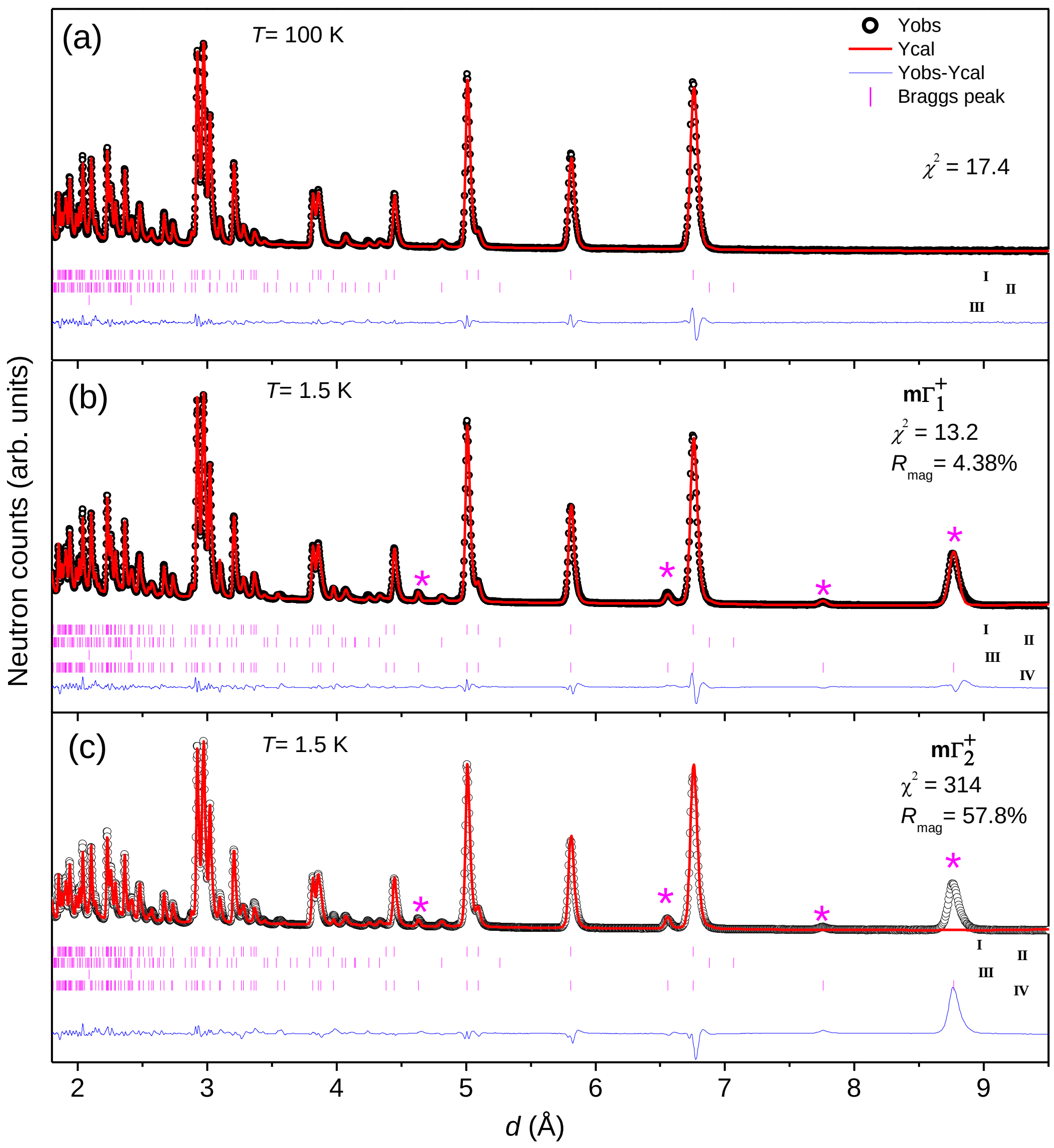

**FIG. 5.** Neutron diffraction patterns of $Li_2Ni_3P_4O_{14}$ measured at (a) 100 K and (b-c) 1.5 K along with refinement by considering magnetic structure as per $m\Gamma 1^+$ and $m\Gamma_2^+$, respectably. Experimental, calculated and difference patterns are shown by black symbols, red curves, and blue curves, respectively. Pink vertical bars are the Bragg peaks positions for nuclear phases of $Li_2Ni_3P_4O_{14}$ (I), $Li_4P_2O_7$ (II), and NiO (III) and magnetic phase of $Li_2Ni_3P_4O_{14}$(IV). The asterisk shows the intense magnetic Braggs peaks of $Li_2Ni_3P_4O_{14}$.

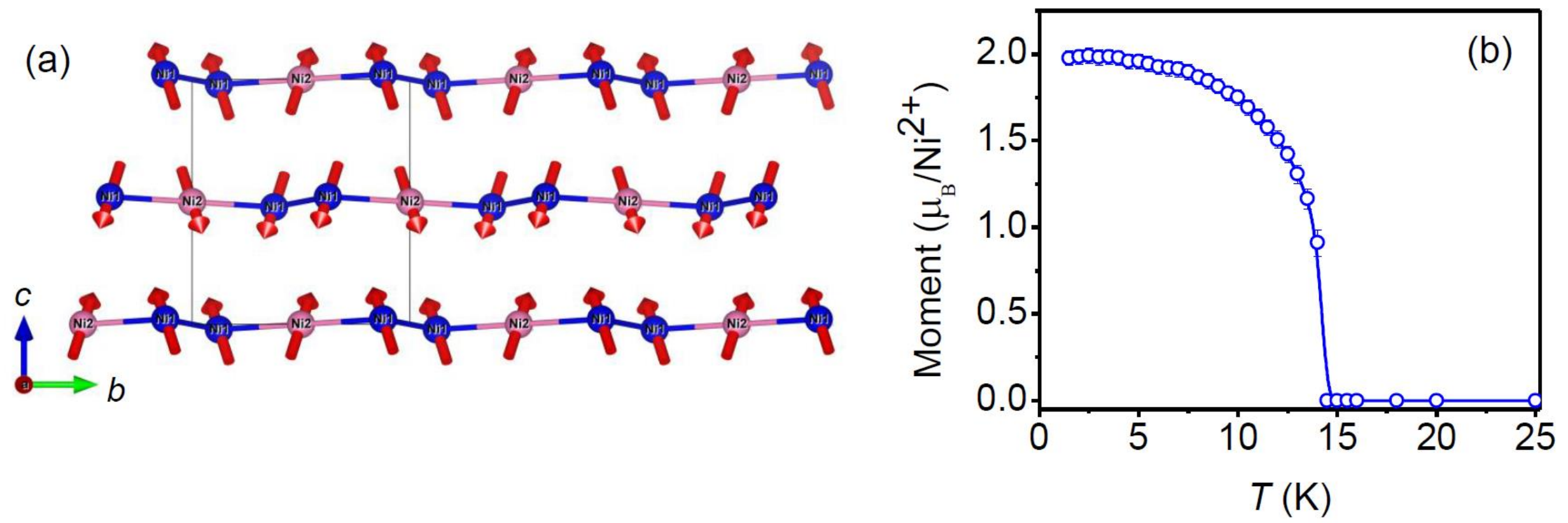

**FIG. 6.** (a) Schematic representation of the derived magnetic structure, characterized by propagation vector $\boldsymbol{k}$ = (0,0,0) and magnetic space group $P21/c.1$. (b) Temperature variation of the ordered magnetic moment derived from Rietveld analysis of the low temperature neutron diffraction patterns.

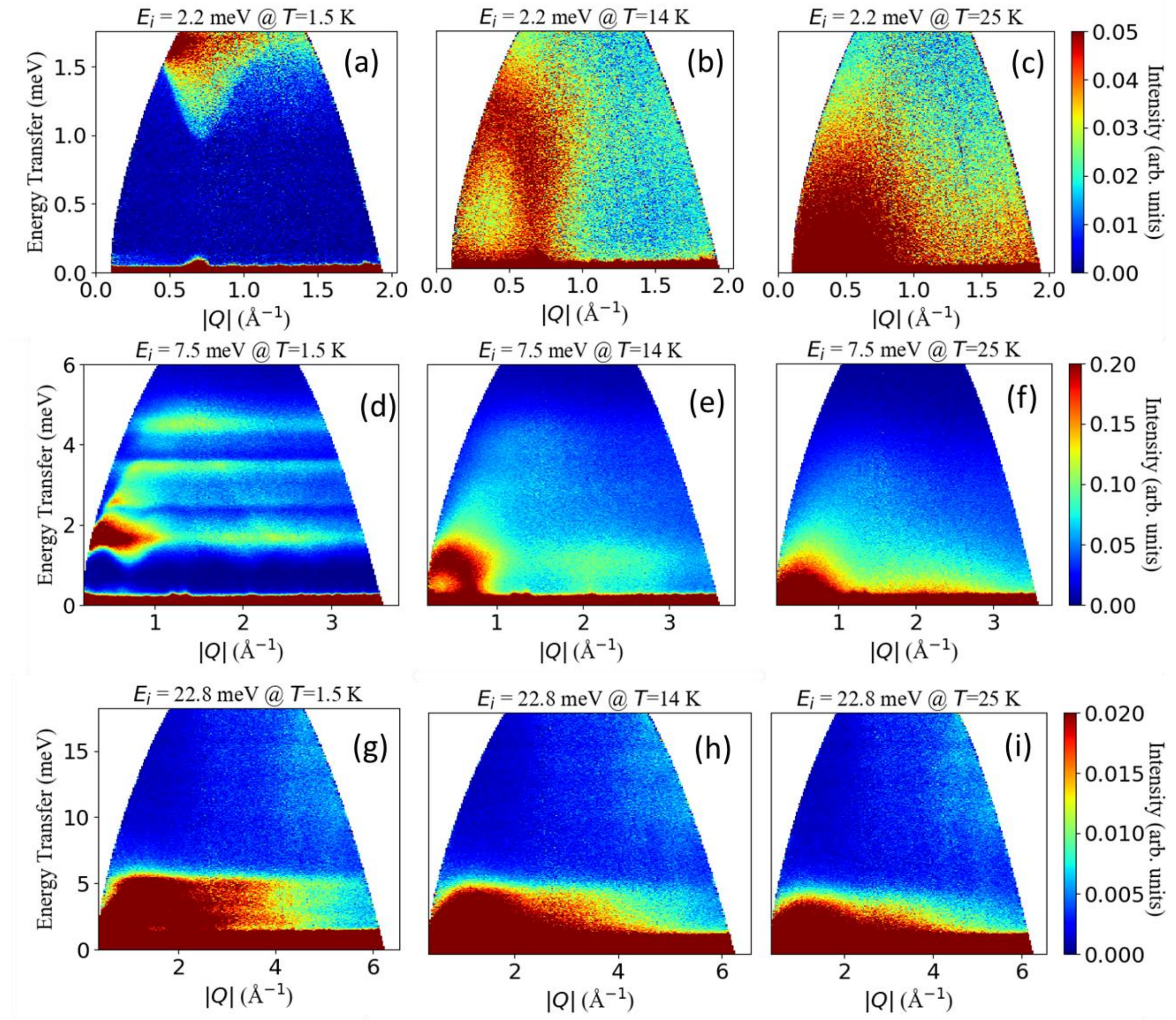

**FIG. 7**. 2D color maps of the INS intensity for $Li_2Ni_3P_4O_{14}$, plotted as a function of energy transfer ($\hbar\omega$) and momentum transfer ($|Q|$). Data were collected on the LET spectrometer at 1.5, 14, and 25 K with incident energies of (a-c) $E_i$ =2.2 meV, (d-f) $E_i$ =7.5 meV, (g-i) $E_i$ =22.8 meV, respectively. The color scales show the scattering intensity $S(|Q|,\omega)$ in arbitrary units.

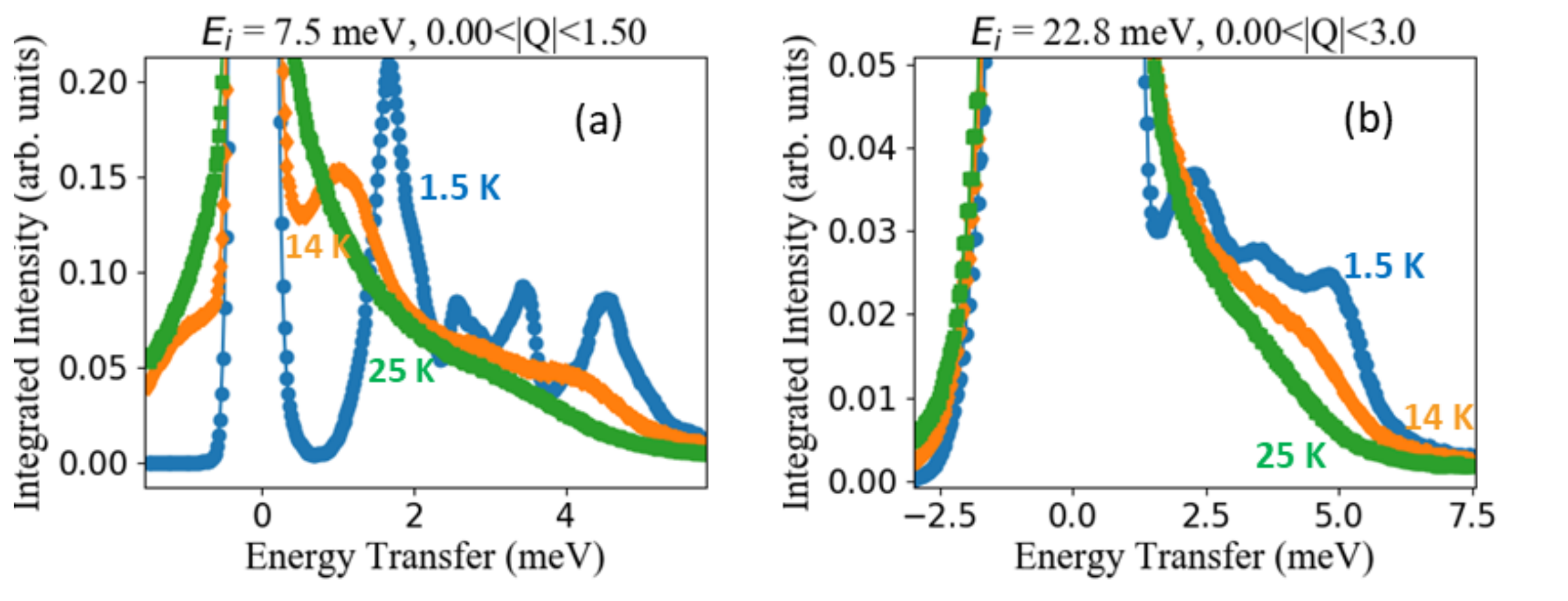

**FIG. 8**. Constant-$Q$ energy cuts of INS data for $Li_2Ni_3P_4O_{14}$, at 1.5, 14, and 25 K, measured with incident energies of (a) $E_i$ = 7.5 and (b) 22.8 meV. The data were integrated over $|Q|$ = 0 – 1.5 Å$^{-1}$ for $E_i$ = 7.5 meV, and over $|Q|$ = 0 – 3.0 Å$^{-1}$ for $E_i$ = 22.8 meV, respectively.

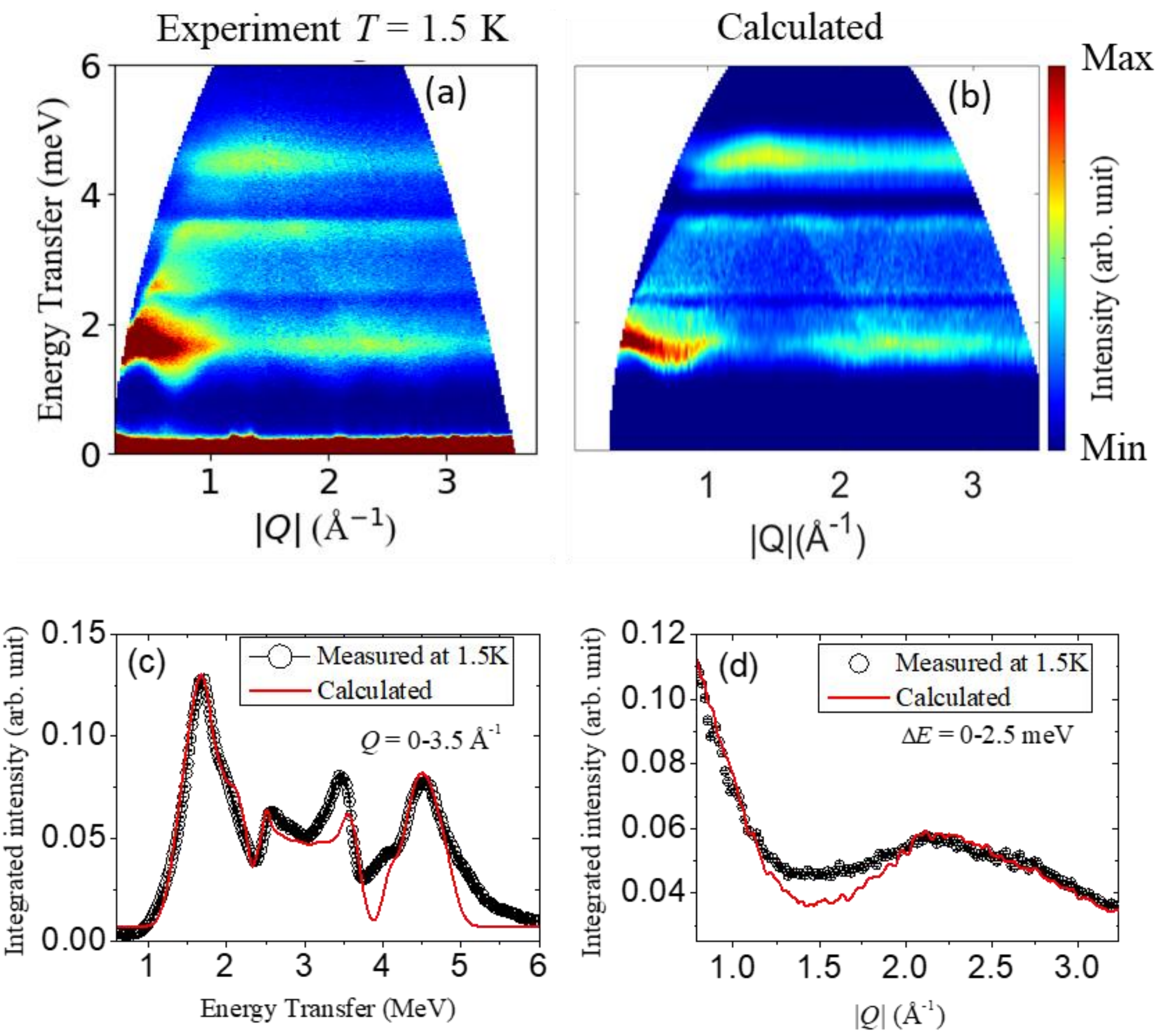

**FIG. 9**. (a) 2D color map of the INS intensity for $Li_2Ni_3P_4O_{14}$, measured at 1.5 K, on the LET spectrometer with incident neutron energy of $E_i$ =7.5meV as a function of energy transfer ($\hbar\omega$) and momentum transfer ($|Q|$). (b) The simulated INS spectrum based on the spin Hamiltonian described in Eq. 1 using the linear spin-wave theory. The color scales show the scattering intensity $S(|Q|,\omega)$ in arbitrary units. (c) and (d) Comparison of constant $|Q|$ and constant energy ($\Delta E$) cuts, respectively, for experimentally measured and calculated INS spectra.

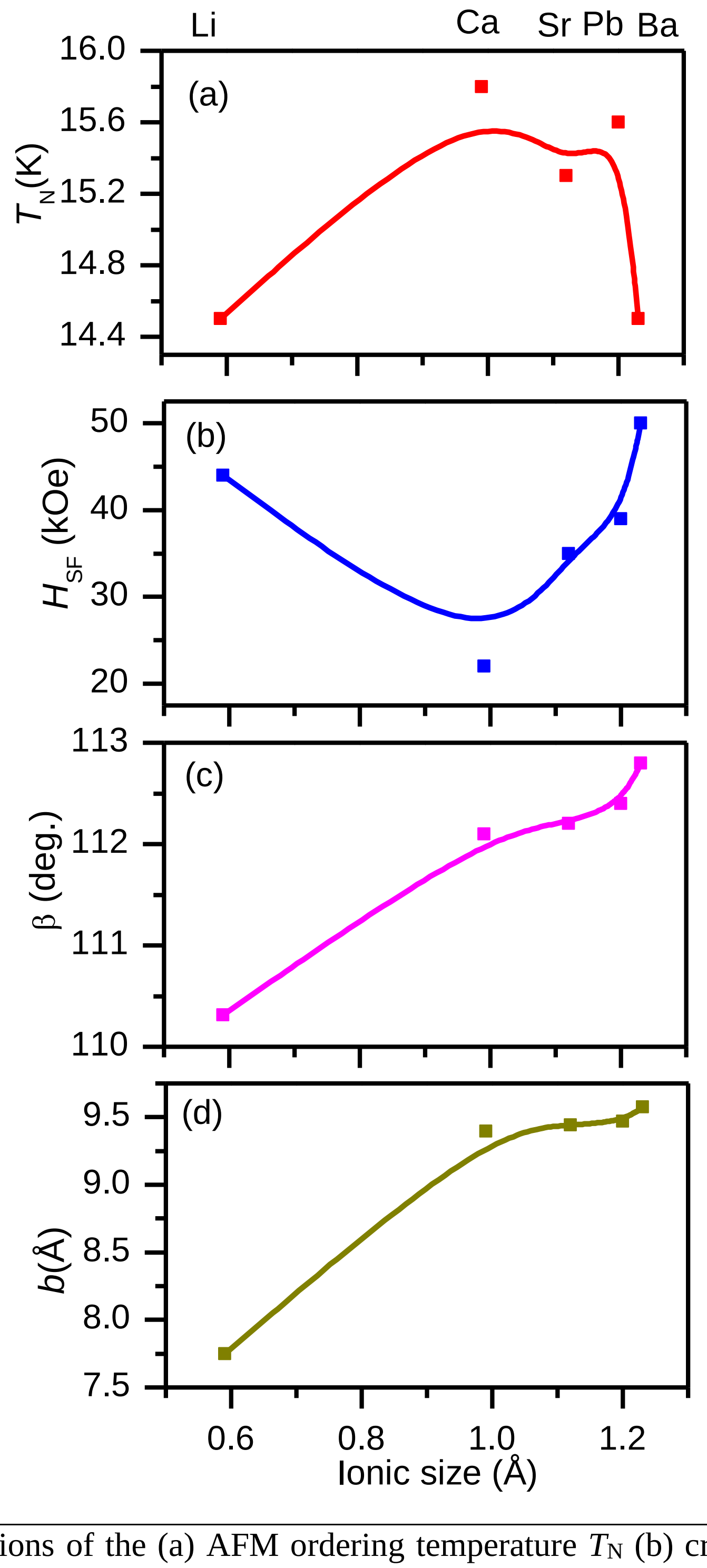

**FIG. 10.** The variations of the (a) AFM ordering temperature $T_N$ (b) critical field $H_{SF}$ for metamagnetic transition, (c) monoclinic angle $\beta$ and (d) lattice parameter *b* with ionic radius of the *A*-site ions in the $ANi_3P_4O_{14}$ compounds. The data for the Ca, Sr, Pb and Ba have been taken from Ref. [21].